# The persistent question of potassium channel permeation mechanisms

Andrei Mironenko[1], Ulrich Zachariae[2], Bert L. de Groot[1], Wojciech Kopec[1]

[1]Computational Biomolecular Dynamics Group, Max Planck Institute for Biophysical Chemistry, 37077 Göttingen, Germany

[2]Computational Biology, School of Life Sciences, University of Dundee, Dundee DD1 5EH, UK

**GRAPHICAL ABSTRACT**

**K$^+$ channel ion permeation mechanisms**

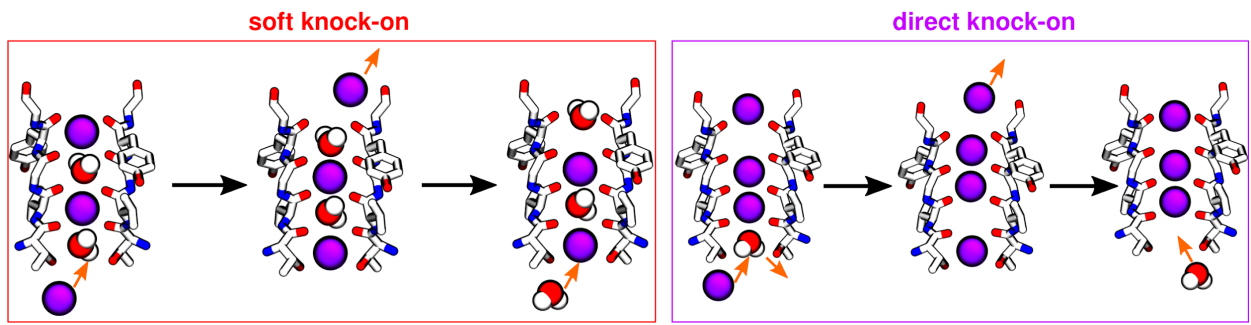

**ABSTRACT**

Potassium channels play critical roles in many physiological processes, providing a selective permeation route for K$^+$ ions in and out of a cell, by employing a carefully designed selectivity filter, evolutionarily conserved from viruses to mammals. The structure of the selectivity filter was determined at atomic resolution by x-ray crystallography, showing a tight coordination of desolvated K$^+$ ions by the channel. However, the molecular mechanism of K$^+$ ions permeation through potassium channels remains unclear, with structural, functional and computational studies often providing conflicting data and interpretations. In this review, we will present the proposed mechanisms, discuss their origins, and will critically assess them against all available data. General properties shared by all potassium channels are introduced first, followed by the introduction of two main mechanisms of ion permeation: soft and direct knock-on. Then, we will discuss critical computational and experimental studies that shaped the field. We will especially focus on molecular dynamics (MD) simulations, that provided mechanistic and energetic aspects of $K^+$ permeation, but at the same time created long-standing controversies. Further challenges and possible solutions are presented as well.

**RESEARCH HIGHLIGHTS**

- $K^+$ channels are characterized by the highly efficient and selective permeation of $K^+$ ions.

- Determining the ion permeation mechanism in $K^+$ channels would explain their remarkable conductive properties.

- Many experimental and computational studies have been performed, and multiple ion permeation mechanisms have been proposed - however, the controversies in the field persist and consensus on which mechanism actually occurs is yet to be reached.

- In this review, we discuss important milestones in research of ion permeation mechanisms in $K^+$ channels and compare the results obtained with different approaches, as well as present further challenges and possible solutions.

## INTRODUCTION

Potassium channels ($K^+$ channels), the most widely distributed ion channels [1], are transmembrane proteins found in almost all organisms [2] and several virus families [3,4]. $K^+$ channels enable the flux of $K^+$ ions across plasma and organelle membranes down the electrochemical gradient.

Potassium channels are characterized by high transport rates (~$10^8$ $K^+$ ions per second, or 1 ion per 10 ns, i.e. near diffusion-limited rates), exquisite $K^+$ selectivity, especially over other monovalent ions such as $Na^+$ (~100-1000 times more selective for $K^+$ than $Na^+$ [5]), and intricate gating mechanisms that allow for current regulation by channel opening and closing, respectively. $K^+$ currents mediated by potassium channels establish a membrane potential, and terminate action potentials in electrically excitable cells such as neurons and cardiac myocytes.

Structurally, a given $K^+$ channel can be divided into a mandatory pore-forming domain and into other optional regulatory domains (Figure 1 A), a voltage sensing or a ligand-binding domain, for example. The overall structure of the pore-forming domain is conserved in all potassium channels, as its primary function is to permeate $K^+$ ions. The regulatory domains are responsible for sensing external stimuli, such as voltage or ligand binding, and allow the channel to gate in response to them, although it is worth noting that $K^+$ channels consisting only of pore-forming domains exist and are capable of gating as well. The variety of regulatory domains and gating mechanisms gives rise to the large size of the potassium channel family [6]. However, the primary function of $K^+$ channels - rapid and selective permeation of potassium ions - is expected to occur in a similar fashion in all $K^+$ channels, and its molecular underpinnings are the main point of this review.

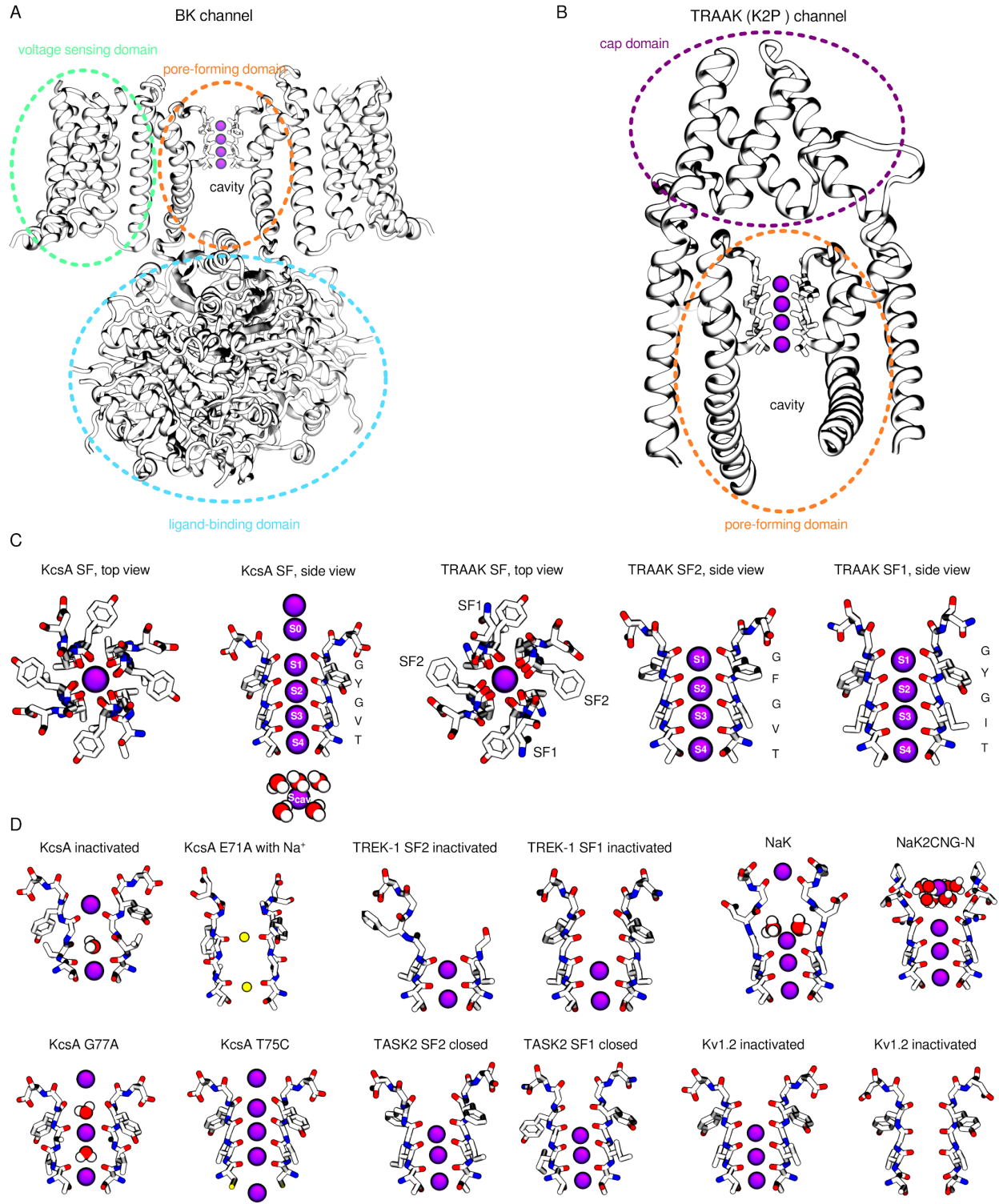

*Figure 1. Overview of potassium channel structures. (A, B) Two examples of potassium channels in open (conductive) conformations - (A) a voltage and calcium-gated big conductance (BK) channel (PDB ID: 5TJ6, two monomeric units out of four are shown for clarity) and (B)*

*TRAAK (KCNK4) K2P channel (PDB ID: 4I9W, parts of the protein blocking the view are not shown). Potassium ions are shown as purple spheres, and the filter forming residues as sticks. (C) Selectivity filter structure and organization. In tetrameric potassium channels, such as KcsA, four identical loops create a 4-fold symmetric filter, with 4 main ion binding sites (S1-S4, PDB ID: 1K4C). In K2P channels, a dimeric organization is observed, with each dimer contributing two loops (SF1 and SF2), resulting in a pseudo-4-fold symmetric filter. (D) Non-conductive conformations of the selectivity filter, or non-selective filters: inactivated KcsA filter (non-conductive, PDB: 1K4D), KcsA filter at 0 mM K+ (non-selective, 'flipped' carbonyl groups and aspartates, PDB: 3OGC, $Na^+$ ions shown as yellow spheres), inactivated TREK-1 filter at 30 mM K+ (non-conductive, part of SF2 not resolved, PDB: 6W7E), filter of the non-selective NaK channel, with two main potassium-binding sites (PDB: 3E8H), filter of the non-selective NaK2CNG-N channel, with three main potassium-binding sites (PDB: 3K06), filter of mutated KcsA at S2/S3 (G77A, PDB: 6NFU), filter of mutated KcsA at S4 (T75C, PDB: 1S5H), filter of the K2P channel TASK-2 (non-conductive, PDB: 6WLV), inactivated Kv1.2 filter (non-conductive, PDB: 5WIE), another inactivated Kv1.2 filter (non-conductive, $K^+$ ions not included in the final model, although visible in the cryoEM map, PDB: 6EBK). Water molecules are shown as red and white spheres.*

**PORE-FORMING DOMAIN AND THE SELECTIVITY FILTER**

The pore-forming domain is conserved in all potassium channels, and in most of them is built up by a tetramer of four identical subunits that surround the central permeation pathway for ions. One class of potassium channels, namely K2P channels (standing for two-pore-domain potassium channels), are dimers, with each dimer providing an equivalent of two units in tetrameric potassium channels, eventually resulting in a very similar pore domain (Figure 1 B). The critical core element of the pore domain is the selectivity filter (SF) - the narrowest part of the pathway, composed of the signature amino acid sequence of $K^+$ channels - typically the TVGYG motif (Figure 1 C); however, certain substitutions are tolerated, such as the Y to F and T to S substitutions in the hERG channel [7]. In tetrameric $K^+$ channels, the signature sequence is identical in all four subunits, whereas K2P tolerates variations within each dimer (see Figure 1 C). Despite these possible variations, the SF, in its conductive conformation, contains four main potassium ion binding sites - denoted S1 to S4 - as well as additional S0 and Scav (cavity) sites above and below the filter. Sites S1 to S3 are formed exclusively by carbonyl-oxygen atoms from the protein backbone of each monomer, whereas site S4 is lined by both carbonyl and hydroxyl oxygens of four threonine residues. As revealed by crystallographic studies, these main binding sites provide a tight coordination for four fully dehydrated potassium ions, replacing water molecules from the ions hydration shell upon ion entry to the SF [8,9]. Both the number of ion binding sites and their geometry are essential for rapid and selective $K^+$ conduction [8–10]. SFs displaying a reduced number of ion binding sites or widened/narrowed SFs usually lose one or both of these key features of ion conduction, indicating ion permeation mechanisms differing from the canonical one, or a complete lack of ion conduction through these filters. Some examples of SF structures that deviate from the typical, conductive conformation are shown in Figure 1D.

Below the SF, the channel forms a central water-filled cavity, where ions remain hydrated. K2P channels further display a unique 'cap' structure above the SF (Figure 1 B).

**$K^+$ PERMEATION MECHANISMS THROUGH POTASSIUM CHANNELS**

The question "how $K^+$ ions pass the selectivity filter", which defines the ion permeation mechanism, is arguably the most fundamental one in all potassium channel research, and not surprisingly has been studied for decades. Predating structural information, electrophysiological measurements between the 1950s-1970s allowed researchers, most prominently Hodgkin, Keynes, Hille, Armstrong, and Bezanilla, to suggest the existence of a potassium-selective channel, in which ions should be constrained to move in a single file, with multiple potassium binding sites [11–13]. The narrowest part of the channel was estimated to be around ~3 Å wide, implicating nearly full dehydration of passing $K^+$ ions due to size constraints. Subsequent development of the patch-clamp recording technique provided insights into single-channel electrical activity.

In the 1990s, studies of several potassium channel mutants led to the identification of pore loops and of the signature amino acid sequence [14,15], culminating in the first structures of the KcsA channel published by MacKinnon's laboratory [16]. Visualising, for the first time, the SF (Figure 1 C) with four potassium ions (S1-S4), quickly led to questions about the number of potassium ions simultaneously bound to the filter, and consequently, about the ion permeation mechanism under electrochemical gradients. Initially, it was assumed that the configuration of four potassium ions lined up in a row would not be stable in the filter, as the distance between the charges would be only ~3.3 Å. This assumption was reinforced by further work by MacKinnon and coworkers, where ion occupancies were obtained through x-ray crystallography by replacing $K^+$ ions with electron-rich $Rb^+$ or $Tl^+$. This study claimed that only two ions occupy the ion binding sites S1 to S4 simultaneously: the ions would reside either in sites S1 and S3 or in sites S2 and S4 (Figure 2, left [17], although for $Tl^+$ the overall number of ions was larger than two). The remaining binding sites were proposed to be occupied by water molecules, to prevent direct repulsive electrostatic interactions between neighbouring ions. The crystal structure of the KcsA channel (PDB ID: 1K4C [18]), with four ion binding sites in the filter, was thus proposed to be a superposition of these two occupancy states of the filter. The two discrete occupancy patterns of the SF: potassium ions in sites S1 and S3, water in sites S2 and S4 (called KWKW or S1,S3), and potassium ions in sites S2 and S4, water in sites S1 and S3 (called WKWK or S2,S4) (Figure 2) immediately suggested an intuitive "knock-on" ion permeation mechanism (later termed also as "soft knock-on" [19]) in the presence of a driving force for ion permeation [20]. In this mechanism, the ion-water pairs would move in concert upon the entry of an additional K⁺ ion from either side of the membrane, leading to a co-permeation of potassium ions and water molecules through the channel, with a ratio of ion-to-water-permeation of 1.

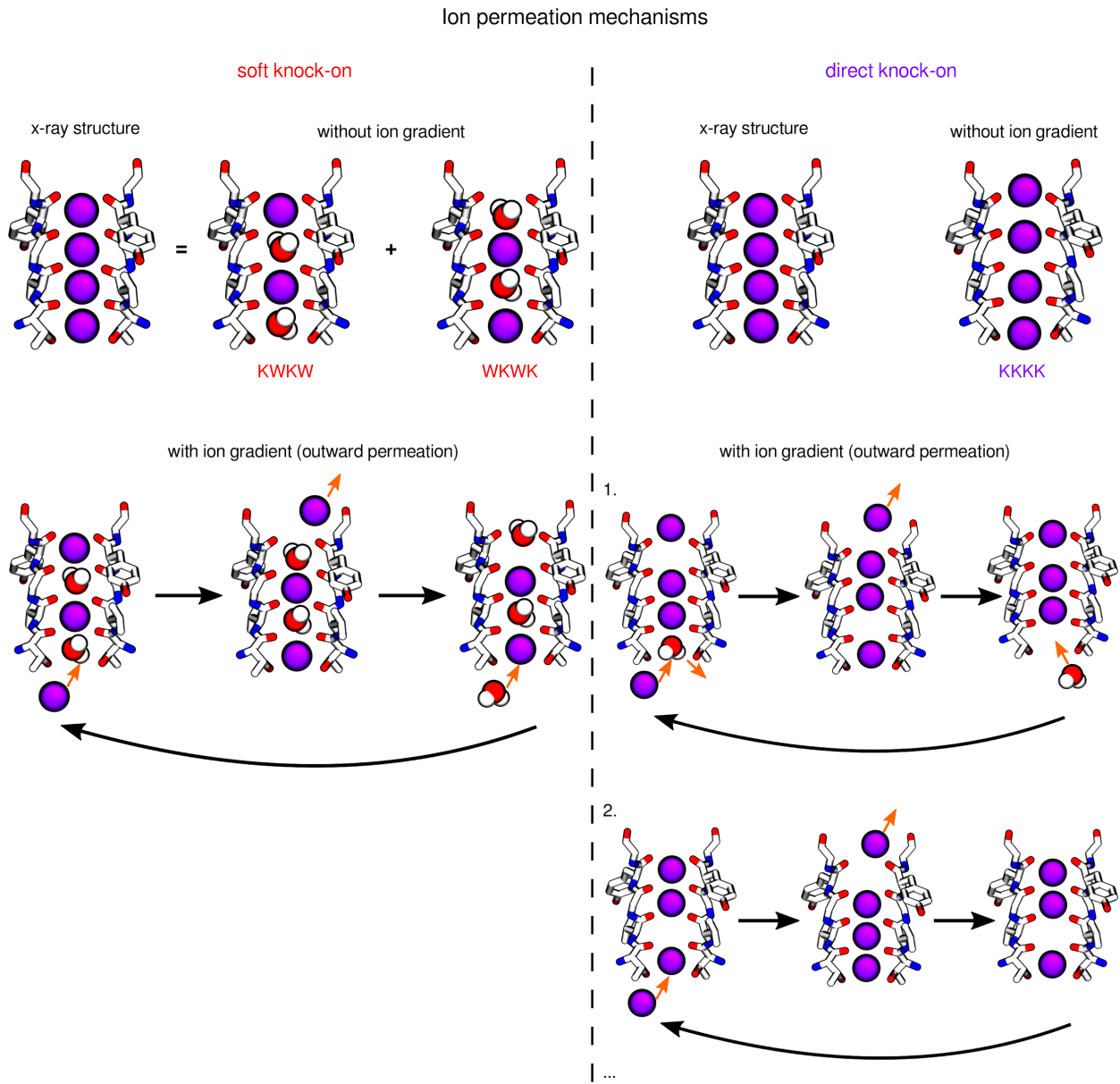

*Figure 2. Overview of proposed ion permeation mechanisms in potassium channels. In the soft knock-on (knock-on) mechanism (left), the occupancy pattern of the selectivity filter seen in the x-ray structure is interpreted as a combination of two isoenergetic, co-existing states, both containing two bound K⁺ ions and two water molecules. With the electrochemical gradient present, outward ion permeation occurs by a shifting between KWKW and WKWK patterns, upon the arrival and binding of a potassium ion and water molecule. Consequently, the KWKW and WKWK states are thought to be predominant also at ion permeation conditions, and one water molecule is postulated to be permeating per one permating/permated K⁺ ion. In contrast, in the direct (hard) knock-on mechanism (right), the SF is postulated to be occupied exclusively*

*by potassium ions (the KKKK occupancy pattern), and at physiological temperatures, a certain dynamics of bound potassium ions is expected. Consequently, ion permeation occurs via short-lived, strong interactions between potassium ions occupying neighbouring ion-binding sites. The water molecules do not play a major role in ion permeation, apart from occasional presence in the water-exposed ion-binding sites (S1 and S4), and do not frequently co-permeate with $K^+$ ions, in contrast to the soft-knock on mechanism. Importantly, several different ion permeation pathways, stemming from various occupancy patterns of the SF, have been observed to date - two possibilities are shown in this Figure; however, several others are possible.*

The soft knock-on mechanism was widely accepted, making its way to several popular biology textbooks, and formed the basis for almost all subsequent studies on ion conduction in potassium channels for at least 10 years [9,21,22]. Early computational studies, which heavily relied on the initial specification of the selectivity filter occupancy, started from the KWKW and/or WKWK patterns (see next section). Similarly, many experimental results - e.g. from electrophysiological recordings [23–25], isothermal titration calorimetry (ITC) [9,21,22], and structures of channels with mutated SF filters [26] - were rationalized and analyzed in the framework proposed by the soft knock-on mechanism.

Subsequent computational studies, however, with longer sampling time and a greater variety of potassium channels, started suggesting a more complicated picture with direct ion-ion contacts (without intervening water molecules) being energetically accessible [27]. In 2014, a combined computational and experimental effort proposed these direct ion-ion contacts as a dominant contribution to ion permeation, coining the term direct knock-on (or hard knock-on [28]). In the direct knock-on mechanism, water molecules do not co-permeate with potassium ions - instead, water is mostly excluded from the SF, apart from an occasional presence in the sites S1 and S4 (Figure 2, right). Consequently, potassium ions are occupying neighbouring ion binding sites, which leads to strong electrostatic repulsion between the ions. The strong repulsion has been proposed to contribute to the high permeation rate of potassium channels [28].

The debate about which of the ion permeation mechanisms predominates under physiological conditions is not, as of now, settled. As we are going to critically discuss in this review, various experimental and computational approaches provide support for either one of the two mechanisms. It is also important to mention that the mechanisms illustrated in Figure 2 present idealized schematic situations. The soft knock-on mechanism, for example, suggests two specific occupancies of the filter to be dominant (KWKW and WKWK) and a strict ion-water co-permeation ratio of 1. Similarly, direct knock-on, although possibly realized through several distinct permeation pathways, imposes an ion-water co-permeation ratio of (or close to) 0. We cannot exclude the existence of mixed mechanisms that can further depend on actual conditions (potassium concentration, the magnitude of membrane voltage or temperature) and the specific potassium channel. Some more exotic mechanisms, involving e.g. even fewer than two $K^+$ ions in the SF at a specific time and increased water fluxes [29,30], have been proposed as well, and will be discussed here accordingly.

**COMPUTATIONAL STUDIES ON PERMEATION MECHANISMS**

As mentioned already, computational studies played a key role in proposing both soft and direct knock-on permeation mechanisms. It is indeed particularly interesting that similar approaches led to contrasting answers. In this part, by careful examination of simulation concepts and protocols, we will show potential reasons for these discrepancies, and provide an overview of the current status in state-of-the-art computer simulations of potassium channels.

*Early studies*

The availability of the first crystal structure [16] of a potassium channel in 1998 - KcsA - greatly expanded the basis provided for computational studies of potassium channels. Using the Poisson-Boltzmann (PB) equation, it was discovered that the cavity and pore helices are electrostatically tuned so that occupancy by monovalent cations is preferred [31]. A seminal early work was the Brownian dynamics (BD) simulation of $K^+$ permeation through a model potassium channel, represented as a transmembrane lumen with several dipoles behind its walls [32]. This study, published in 1999, was essentially the first simulation of $K^+$ current through a potassium channel. These simulations predicted that a potassium channel is simultaneously occupied by two $K^+$, with the ion entering the channel and escaping the potential well created by the mouth dipoles as the rate-limiting step. The study showed good agreement with experimental data in terms of conductance and the concentration-conductance relationship. However, due to the simplicity of the model, and the fact that water was modelled implicitly, one should treat these results with caution.

The first molecular dynamics studies of the KcsA channel focused on the dynamics of water and potassium in the SF. They showed spontaneous transitions between several SF configurations. In particular, the transition of potassium ions from S1,S3 to S2,S4 occupancies was observed [33,34], as well as a concerted transition of a configuration with $K^+$ in Scav, S3, S1 to S4, S2, S0, with water molecules interspersed between the $K^+$ ions [35]. These simulations also provided insight into the basis for $K^+$ selectivity over $Na^+$ - as the fluctuations of the SF were larger than the difference between the radii of $K^+$ and $Na^+$ (the root-mean-square fluctuations were on the order of 1 Å), so the rigidity of the SF was deemed not to explain the ion selectivity of KcsA [35,36].

Despite their success, the limited computational power at the time precluded the observation of complete permeation events using unbiased MD simulations of KcsA. Instead, most of the early studies were performed using free-energy calculations based on predefined ion positions. These comprise two main methods: 1) alchemical free energy calculations or 2) potential of mean force (PMF) based calculations. In alchemical free energy calculations [37], a given part of the system (e.g. a K$^+$ ion) is transformed into a different one (e.g. a water molecule) via a non-physical path. Then, the free energy difference between these states is calculated, e.g. using the Zwanzig equation [38], when the free energy perturbation method (FEP) is used. In PMF calculations, a biasing potential is used to increase sampling of the phase space. For instance, in the umbrella sampling method [39] a path along chosen coordinates (e.g. the position of ions along the channel's pore axis) is divided into a series of overlapping 'windows', which are sampled separately in MD simulations, and, finally, the free energy distributions are 'glued' together to yield a complete energy surface along the coordinates of interest (or 'PMF').

The seminal work by Åqvist and Luzhkov [40], in which the relative free energies of multiple ion/water configurations in the SF were studied using FEP - with 1,2,3 or 4 K$^+$ in the filter, and the remaining sites occupied by water molecules - proposed a preference for two potassium ions in the filter - either in S2,S4 or in S1,S3 occupancies. Another key study was carried out by Berneche & Roux [41], where umbrella sampling was used to calculate the free energy (PMF) surface for positions of three ions in the SF along the pore axis. The study suggest that permeation in KcsA can occur via two pathways: The first pathway, reminiscent of the 'knock-on' mechanism, originally proposed by Hodgkin and Keynes [11], starts from a configuration with K$^+$ in S1, S3, separated by a single water molecule. An ion in the cavity then approached the intracellular entrance to SF, pushing the ion pair to S2,S4, resulting in the exit of the outermost ion to the extracellular side. Subsequently, the two ions remaining in SF were able to move back and forth in the SF overcoming a small barrier of 1 kcal mol$^{-1}$. The largest free energy barrier calculated for this pathway was on the order of 2-3 kcal mol$^{-1}$, and thus the process was essentially diffusion-limited. The second pathway comprises the ion pair in the SF moving to the extracellular side, thus leaving a vacancy on the intracellular side of the SF that is subsequently filled by an incoming ion. The barrier in this case was 3-4 kcal mol$^{-1}$ high. Importantly, both pathways were essentially transitions between states with two and three potassium ions in S1,S3 and S0,S2,S4 occupancies, respectively. Each ion pair was separated by a single water molecule, such that both pathways are variations of the soft knock-on mechanism (Figure 2). The existence of two distinct pathways was proposed to complement the small free energy barriers, thus leading to the high conductivity of K$^+$ channels.

Based on the all-atom PMF obtained by Berneche & Roux, BD simulations of ion permeation in KcsA under transmembrane voltage were performed [42]. Permeation occurred through the dominant pathway in their PMF calculations, that is with potassium ions occupying [S3, S1] -> [S4, S3, S1] -> [S4, S2, S0] -> [S4, S2] -> [S3, S1] for outward, and in the opposite direction for inward permeation, with water between K$^+$. This model's predictions were in broad agreement with experiment [5], thus further solidifying the ground for the knock-on mechanism, with water co-permeation.

However, later, based on umbrella sampling calculations by Furini and Domene [27], it was revealed that other K$^+$ permeation pathways are also possible. Simulations were initiated in three different SF configurations - in one of them, K+ ions were separated by water molecules (KWKW, Figure 2), but in two other configurations, either vacancies or direct contacts between ions were considered. Moreover, the simulations were carried out for two different K$^+$ channels, namely KcsA and KirBac. Two permeation pathways were observed - the previously described water-mediated knock-on [41], as well as a novel, water-free permeation mechanism which involved formation of direct ion-ion contacts. Importantly, the free energy barriers along these two mechanisms were essentially identical - on the order of 2-3 kcal mol$^{-1}$ for outward permeation. Moreover, as revealed by FEP calculations, the energy cost to arrive from a given 'water-mediated' configuration of the SF to an analogous 'water-free' configuration, was rather

low - less than 2 kcal mol$^{-1}$, indicating a similar plausibility of both pathways.

Importantly, in both seminal PMF-based studies of ion permeation in K+ channels by Berneche & Roux [41,42] and Furini & Domene [27], the authors used modified Lennard-Jones (LJ) parameters describing interactions between K+ ions and carbonyl oxygens (termed the NBFIX correction). We will discuss this point later in this review.

*Direct simulations of permeation events*

In 2006, Khalili-Araghi *et al.* described the first direct molecular dynamics simulation of ion permeation in a K$^+$ channel [43]. For the Kv1.2 channel, using its then-recently determined crystal structure in the open state (PDB ID: 2A79) [44], they observed several permeation events in 25 ns-long MD simulations with applied electric field, corresponding to a transmembrane voltage of 1 V. Permeations occurred via the water mediated knock-on mechanism, in agreement with earlier studies [41,42]. However, such high voltages might have affected the conduction mechanism, which was acknowledged by the authors. In addition, to ensure SF stability, its backbone dihedral angles were restrained to the values extracted from the x-ray structure, possibly affecting the dynamics of the channel. The possible issues of using high voltages, as well as effects of the SF structure on the permeation mechanism will be discussed in more detail below.

With advances in computing power and development of specialized hardware, observing multiple permeation events on a microsecond timescale in MD simulations of potassium channels became possible. In 2009, the Shaw group published a study about the permeation of about 500 ions through the Kv1.2 pore domain (PDB IDs: 2A79 [44], 2R9R [45]), over the course of ~30 µs with applied electric field [46]. The soft knock-on permeation mechanism was observed. Potassium ions showed a preference for binding at sites S2,S4 over S1,S3, with water having the opposite preference, and the formation of the knock-on intermediate with two ions forming a direct contact - i.e. the [S6, S4, S2] to [S5, S4, S2] transition - was the rate-limiting step (in this work, S5 and S6 denote two subsequent ion binding positions below the SF, along its axis). The voltage range was reported as -180 mV to 180 mV. However, in the method of computing voltage, $V = E\Delta z$, where $E$ denotes the electric field, $\Delta z$ was taken as the length of the SF, instead of the length of the simulation box along the membrane normal, which produced underestimated voltages, obscuring the interpretation of these results. The issue was further clarified by Gumbart *et al.*, who rigorously showed that the membrane voltage in periodic systems should be calculated with $\Delta z$ describing the length of the box in the direction of the membrane normal [47].

In a follow-up work, the Shaw group thus repeated the simulations of Kv1.2 with the desired voltage, as well as performed similar simulations of the KcsA channel [48]. For Kv1.2, the simulated conductance at around 300 mV was much lower (~2 pS) than the experimentally determined one (73 pS). One of the possible reasons for this discrepancy considered by the authors was an overestimated interaction energy of $K^+$ ions with the SF carbonyl groups in the used CHARMM force field (inspired by earlier PMF-based works); surprisingly, however, reducing the corresponding LJ interactions via the NBFIX correction did not increase the conductance. On the other hand, this change revealed the dependence of the permeation mechanism on the applied force field: weakening the interactions in Kv1.2 changed the mechanism from the water-mediated knock-on to a different one, where the SF occupancy by $K^+$ ions decreased from ~2.6 to <2, due to $K^+$ ions dissociating more readily from S0 and S5. In addition, during $K^+$ permeation, the knock-on intermediate [S5, S4, S2] was bypassed - the [S5, S3, S1] state emerged directly from [S6, S4, S2]. In simulations of the constitutively open KcsA with reduced $K^+$–SF interactions, water-mediated knock-on occurred, similar to the mechanism observed by Berneche & Roux [41,42,49], whereas using default LJ interactions led to a different knock-on mechanism with a predominant three-ion state, [S4, S2, S1]. In this pathway, the SF occupancy by K$^+$ ions was increased to ~3.3, and no water co-permeation was observed.

In 2014, an alternative method for generating a transmembrane voltage was applied for the first time to study potassium channels [28] computational electrophysiology (CompEL). Instead of applying an electric field, in CompEL the system is separated into two compartments by a double membrane setup, and a charge imbalance is created in each compartment, thus directly simulating an electrochemical gradient [50,51]. This charge imbalance, calibrated to correspond to a predefined transmembrane voltage across both membranes, is maintained during the simulation by monitoring ion permeation events and swapping ions in solution with water molecules between the two compartments. Compared to an applied electric field, CompEL, despite an increase of the system size after adding a second lipid bilayer, is only slightly more computationally expensive, due to the possibility to insert an ion channel into the second bilayer, that effectively doubles the sampling (for detailed discussion see Kutzner *et al.* [51]).

In our simulations of the bacterial MthK potassium channel, we addressed the question whether the method of generating a transmembrane voltage might affect the permeation mechanism [52]. It turned out to not be the case - both methods (CompEL and applied field) yielded similar currents in a moderate voltage range (200-300 mV), dominated by the direct knock-on mechanism. In certain cases, however, the applied external electric field method may lead to artifacts. For example, the relation between the applied electric field and the locally acting field being unknown *a priori* [53] - it depends on the electrostatics scheme used (cut-off, reaction field, Ewald summation), periodic boundary conditions (PBC), and on the dielectric distribution across the simulation box. It has been shown, however, that the combination of the commonly used PME electrostatics scheme, a 'tinfoil' PBC, and single lipid bilayer follow $V = E\Delta z$, and only large fluctuations of the membrane dielectric properties (e.g. in electroporation [54]) might cause deviations from this formula.

Using CompEL, currents through KcsA, Kv1.2, and MthK were simulated at near-physiological voltages, with a total of more than 1300 permeation events over the course of ~50 μs [28]. In all three channels, permeation occurred via a water-free mechanism under formation of direct ion-ion contacts. At the core of this mechanism lies a concerted transition of $K^+$ ions from [Scav, S3, S2] to an intermediate [S4, S3, S2], with a following 'knock-on' of the central ion pair to S2,S1, and, finally, restoration of the central ion pair at S3,S2, via the exit of $K^+$ ions at S1 to S0, and a shift of S4 to S3.

Other similar pathways are also possible, united by the lack of water co-permeation: water molecules were found to frequently occupy the S1 and S4 sites but not the central binding sites and, therefore, did not co-permeate (Figure 2, right panel). The rate of $K^+$ ions leaving the extracellular site was found to be determined by the rate with which incoming intracellular ions arrive at the SF, thus the model inherently implies that $K^+$ ion permeation is diffusion-limited as long as $K^+$ ions occupy S2 and S3. This mechanism was observed independently of the force field and water model used (see below). Only supraphysiological voltages (~1 V) led to the co-permeation of water, although position restraints had to be applied to the SF to preserve its structure under these conditions. The conductance observed was in agreement with experimental results (up to a factor of ~2, similar to the experimental range of variations) [5].

The direct knock-on mechanism was subsequently consistently observed for several potassium channels and a range of different simulation conditions [52,55–58]. Also, ion permeation in some K2P channels (TRAAK, TREK-1, TREK-2) occurred via this mechanism [57–59]. In Kopec *et al.* [52], the origins of the exquisite $K^+$/$Na^+$ selectivity was studied in KcsA, MthK, Kv1.2 W362Y and NaK2K channels. Binding of $Na^+$ drastically reduced permeation in all $K^+$ channels studied, and substantial Na$^+$ currents (albeit an order of magnitude lower than maximum K$^+$ currents) were observed only in KcsA and NaK2K in the complete absence of K$^+$, in accordance with experiments and the known robust selectivity of K$^+$ channels. The direct knock-on mechanism suggests that the complete desolvation of ions underpins selectivity, because the desolvation penalty for K$^+$ is much smaller than for Na$^+$. Indeed, in this study [52] the ion selectivity of K$^+$ channels, as well as permeation rates, were found to be drastically diminished with any level of water co-permeation. The water-free ion permeation was found to be in agreement with recalculated IR spectra of ion-occupied KcsA SFs (see experimental section), as well as measured ion interaction energies [60], whereas the calculated Na$^+$/K$^+$ permeability ratio of 0.02-0.04 is in good agreement with the experimental values of 0.006-0.4 [5,61,62].

Of particular interest here are the MD simulations of the pore domain of the Kv1.2 channel, initiated from the 2.4 Å crystal structure of the Kv1.2-Kv2.1 paddle chimera channel (PDB ID: 2R9R). As described above, previous long simulations on specialized hardware showed the soft knock-on mechanism in this channel. The electrostatic profile of the Kv1.2 SF was however markedly different from the one for KcsA, which was explained by different orientations of conserved aspartate residues behind the SF (Figure 1 C). Moving the side chains of these residues (D80) in KcsA from configurations initially oriented towards the SF center to an extracellular-facing orientation ('flipped' orientation, that was observed experimentally for the KcsA E71A mutant at low potassium concentration) drew the profile closer to that of Kv1.2 [48]. Indeed, in Kv1.2, the corresponding D375 residues were almost exclusively found in the "flipped" orientation. As shown in Kopec *et al.* [52], such "flipped" aspartates in Kv1.2 lead to an unstable SF, allowing water to enter the SF, and subsequently reducing both potassium current and selectivity, what is in good agreement with experimental studies of channels modified at the corresponding aspartates [63,64]. Similar side chain flips in the corresponding N147 and D256 residues were later observed in simulations of the TREK1 channel of the K2P family, and were associated with the loss of conductivity [58]. For Kv1.2, Kopec *et al*. proposed the W362Y mutation, that strengthens the hydrogen bond to D375 and reduces the flipping rate, or the use of position restraints on D375 to keep it in a non-flipped orientation. Both approaches resulted in reduced water-permeation and increased direct ion-ion contacts, which in turn increased the current and selectivity to the levels observed for other channels in this study [48]. These observations strongly suggest that the "flipped" aspartate conformation, attained in e.g. unrestrained MD simulations of Kv1.2 starting from PDB ID 2R9R, is likely not representative for the conductive SF.

Alongside the two mechanisms described above - the water-mediated and the direct knock-on mechanism - alternative mechanisms of ion permeation were observed in MD simulations as well. Recently, an uncommon permeation mechanism was revealed for the hERG1 channel by Miranda *et al.* [65]. There, a very low outward conductance of ca. 1 pS, in reasonable agreement with the 3-5 pS flickering conductance from experiments [66–70] was accompanied by low stability and high distortion of the SF. The SF was occupied by only ~1 $K^+$ ion, and rare permeation events were water-mediated. Potential reasons for this behavior might be: firstly, the starting CryoEM structure, which is of relatively low resolution (3.8 A, PDB ID 5VA2 [7]), coupled with the absence of water molecules behind the SF, usually observed in high-resolution crystal structures of potassium channels (see Figure 3 and the discussion in the NMR part). Secondly, the voltage used in the simulations was relatively high - 500 and 750 mV. Finally, the sequence of the SF of hERG is unusual, namely $^{624}$SVGFGN$^{629}$, with residues S624 and N629 differing from the more common T and D, respectively (N629 replaces the aspartates discussed above for Kv1.2; the corresponding T to S mutation in MthK drastically reduces both the open probability and the single-channel conductance [71]). It is important to note that non-selective $K^+$ channels such as NaK [10] and NaK2CNG-N [52] are also characterised by a distorted and water-permeating SF, and therefore the question of mechanism of ion permeation and selectivity of such SFs should be addressed in future studies. Another example of a non-traditional mechanism was observed in work by Sumikama & Oiki [29,30], where Kv1.2 [29] displayed a mechanism reminiscent of the vacancy model proposed earlier [72]. Here, an ion pair in S3, S1, with a water molecule in between, shifts to S2,S0, leaving a vacancy at S3 behind, which is then filled by a water molecule from S4 upon entry of $K^+$ to this site. In their studies, KcsA [30] showed two modes of permeation: one of which resembled a water-mediated knock-on, and another consisting of the spontaneous exit of a $K^+$ from a 2-$K^+$-occupied SF, followed by another ion entering the SF from the cavity. It should be noted, however, that both the simulations of the hERG1 channel [65] and Sumikama & Oiki's work [29,30] on Kv1.2 and KcsA featured supraphysiological voltages, which may have affected the channels, and specifically, the SF structure [28]. In turn, this effect might have led to alterations in permeation mechanisms. In conclusion, these observations indicate a potential variability of permeation mechanisms for channels with different structures, as well as for different simulation conditions.

*Computational studies of $K^+$ channels: technical aspects*

The above mentioned observations bring us to the discussion concerning simulation parameters and conditions one might use to obtain reliable results in MD simulations of ion permeation in potassium channels. One of the critical aspects in any MD study is the choice of the force field. In this section, we will focus on the two force field families that are most frequently used in simulations of ion channels - AMBER and CHARMM, will discuss the parameters, as well as how they affect permeation and structure of ion channels.

To properly characterize ion permeation, an accurate description of ion-water and ion-protein interactions is critical. To accurately tune the parameters of ion-water interactions, several optimization strategies are possible, such as reproducing ion hydration free energies, ion radii, structure properties of salt crystals, and others [73–75]. Single-ion hydration (solvation) free energy is a common parameterization target. However, its experimental measurement is not straightforward - in contrast to neutral salts, for charged moieties there is no thermodynamic system to ambiguously measure hydration free energy, and additional assumptions are necessary (this also holds true for $K^+$/N-methylacetamide solvation free energies discussed below). A thorough overview of approaches and caveats in the field of single-ion solvation are presented in the book of Hünenberger & Reif [76]. Potential complications involve (but are not limited to): first, the existence of several target (experimental) single-ion hydration free energies - e.g. those of Burgess [77], Marcus [78], Schmid [79], Tissandier [80], and others. Second, the method of computing free energies might affect the final result if specialized corrections are not applied [81,82]. Third, it has not always been clear if the contribution from the air (vacuum)-liquid interface should be included in parametrization calculations or not.

Practically all MD studies of $K^+$ channels that used the CHARMM force field (with the CHARMM-specific TIP3P water model) employed the default CHARMM ion-water interaction parameters derived by Beglov and Roux [83], targeting Burgess hydration free energy, without any additional corrections. In simulations of several potassium channels, these parameters led to water-free ion permeation *via* direct knock-on [28,52,55,84,85]. A notable issue has been the parameterization of ion-water interactions in AMBER force fields. The default AMBER ff9X force fields have used the Åqvist parameters for monovalent cations [86] mixed with Dang monovalent anion parameters [87]. This combination leads to crystal salt precipitation well below the saturation limit, as well as to an underestimated stability of certain DNA structures [88]. Despite the acknowledgement of these issues, they were not systematically addressed for over a decade until 2008, when Joung & Cheatham published three sets of optimized ion parameters for monovalent ions. They aimed to reproduce Schmid hydration free energies in conjunction with three water models (TIP3P, TIP4P$_{EW}$, and SPC/E), as well as lattice energies and lattice constants of salt crystals [73]. These newer parameters have been used in our simulations of potassium channels with the AMBER force field family (AMBER ff99sb or ff14sb) [28,52], as well as in other simulations of KcsA, TRAAK [84,85] and Kir channels [89,90]. In all of these studies, the direct knock-on mechanism was observed. Of particular interest, in Kopec *et al.* [52], we have verified for the KcsA channel that three water models TIP3P, TIP4P$_{EW}$ and SPC/E combined with respective Joung & Cheatham ion parameters do not affect ion permeation characteristics. In contrast, Sumikama & Oiki simulated KcsA [30] and Kv1.2 [29] channels using the older AMBER ff94 force field, together with Dang ion parameters, and observed water-mediated ion permeation. However, these studies differ in other conditions as well, such as high (ca. 800 mV) membrane voltages and the water presence in starting structures. Therefore the exact effect of ion parameters on ion permeation in the AMBER force fields is still unknown.

Ion-protein interactions represent another important aspect of force field parameterization. A controversy in the field of MD simulations of potassium channels has been the use of force fields with less attractive Lennard-Jones parameters for ion-backbone carbonyl interactions through the so-called 'NBFIX' correction (historically, potassium channel simulations were among the first ones with such modified interactions, thus the correction was often called simply 'NBFIX'. Since then, however, the approach gained popularity for fine-tuning many other Lennard-Jones parameters [91]. In this review, we use the 'NBFIX' term exclusively when referring to modified potassium ion-backbone carbonyl interactions). Conductance and selectivity are largely governed by these interactions, therefore several studies aimed at improving them. A popular approach involves using the energetics and thermodynamics of $K^+$ interaction with N-methylacetamide (NMA) - a model for the protein backbone carbonyls. Initially, in 2002, it was postulated that in traditional force fields, e.g. CHARMM22 and its further iterations (with default CHARMM $K^+$ parameters [83]), $K^+$ binds too strongly to the SF carbonyls, and that introducing a NBFIX correction weakening, these interactions would lead to a more accurate representation of $K^+$/NMA solvation free energies [36,92]. However, the experimental solvation free energy of $K^+$ in NMA was not known at that time. In 2010, a combined experimental and computational effort led to the determination of the KCl solvation free energy in NMA [93]. The individual contributions from $K^+$ and $Cl^-$ were then approached computationally, due to the aforementioned complications with single-ion solvation free energy measurements [76]. Based on $K^+$/NMA binding energies obtained via *ab initio* quantum mechanical calculations in the gas-phase, the newly developed Drude polarizable force field provided individual solvation free energies for $K^+$ and $Cl^-$. Adding them together reproduced the experimental value of KCl, thus strongly supporting the obtained values. Of particular interest, the fixed charge CHARMM27 force field with original LJ parameters (without NBFIX) between $K^+$ and NMA, reproduced $K^+$/NMA solvation energies better than the NBFIX correction, when compared to the Drude values. Despite that, simulations using NBFIX appeared in further studies [23,94].

An interesting example of force field dependence was demonstrated in the PMF-based calculations by Heer *et al.* [94]. In the seminal work by Berneche & Roux [41] that provided an energetic basis for the soft knock-on mechanism, a KcsA structure with a closed intracellular gate was used (PDB ID: 1BL8 [13]) for PMF calculations. Heer *et al.* postulated that such structures possess a SF with high ion binding affinity and high barriers for ion permeation, i.e. are non-conductive, due to the known allosteric communication between SF and transmembrane helices [94]. The opening of the inner gate by the outward motion of these helices switched the SF conformation to a conductive one. A conductive SF was characterized by relatively high structural fluctuations. The main difference between these two studies (apart from a different structural model of KcsA) was the version of the CHARMM force field. The CHARMM22 force field was used by Berneche & Roux [41], which, according to Heer *et al.*, overestimated the flexibility of the protein main chain, thus hindering this allosteric modulation and gating by the SF itself. The use of CHARMM36 with corrected terms for backbone dihedral angles in Heer *et al.* consequently allowed to distinguish the conductive and non-conductive states of the SF [94]. A full interpretation of this study is however difficult due to initial occupancies of the SF that included states characteristic for soft knock-on only and the use of NBFIX. Nevertheless, an important implication derived from this study is that the ion binding affinity is strongly affected by the functional state of the SF. We followed this idea and showed current variations as a function of the SF width in the MthK channel [55]. For both AMBER ff14sb (with TIP3P water and Joung and Cheatham ion parameters) and CHARMM36 (without NBFIX) force fields, the currents around the physiological opening of the MthK gate were dominated by the direct knock-on mechanism.

However, some variations due to the force field family were observed, which is expected because of the different energetics of potassium-backbone carbonyl interactions. For instance, recent work by Ocello *et al.* [85] on the TRAAK channel observed that a combination of the AMBER ff14sb force field and an SF initially occupied by $K^+$ only is required to observe substantial conductance (on the order of 2-9 pS) at 100-200 mV. Water between $K^+$ ions led to filter instabilities producing reduced currents, and permeation events were possible only after water was expelled from the SF. Switching to the CHARMM36m force field reduced the probability of finding the SF in a native conformation by half, with water-mediated configurations being non-permeable (stability issues were observed for KcsA [84] with CHARMM36m as well).

Gating of TRAAK through bending of transmembrane helices around a conserved G268 residue was also affected by the force-field choice, with AMBER producing a better preserved initial conductive state of the channel, whereas using CHARMM transitions between conductive and non-conductive states were observed more frequently.

As mentioned above, the structure of $K^+$ channels - and the SF in particular - also affects their permeation properties. Distortions in the SF (e.g. flipping of SF residues, SF widening and/or collapse) caused e.g. by high voltages [28] and/or simulation artifacts due to the low-resolution initial structures with no water molecules behind the SF, may explain the unusual mechanisms observed in such conditions [29,30]. Notably, high voltages in general tend to promote water-mediated conduction mechanisms [28,52,65], whereas the direct knock-on was consistently observed even at low voltages - in works by Kopec *et al.*, [55] (150 mV, MthK), Ocello et al., [85] (100 mV, TRAAK), Lolicato *et al.*, [58] (40 mV, TREK-1) - further suggesting an impact of voltage on permeation mechanism.

To summarize, despite more than two decades of computational studies of $K^+$ channels, a number of controversies and discrepancies between different studies still exist. We were able to trace down several causes of such discrepancies, such as initial structures and/or force field parameters as well as the strength of the applied electric field. However, it is encouragingthat recent simulations of several potassium channels, carried out by various independent research groups and using modern force fields, are mostly in broad agreement. These simulations predominantly show water-free ion permeation enabled via direct ion-ion contacts [52,55,58,84,85]. However, MD simulations of $K^+$ channels still systematically produce currents below experimental values [25,85]. One of the likely reasons for this discrepancy is the use of force fields that do not explicitly account for induced electronic polarization [85,92], and therefore, the development of polarizable force fields for K$^+$ channels represents an important area of future research.

**EXPERIMENTAL STUDIES ON PERMEATION MECHANISMS**

*X-ray crystallography*

As mentioned in the introduction, a careful analysis of ion occupancy in the crystal structure of KcsA, aided by anomalous diffraction of thallium ions, a mimic for potassium, led to the interpretation of alternating ion/water occupancy in the selectivity filter [17]. Supported by electrophysiological streaming potential measurements (see next section), that imply ion/water co-permeation in a one-to-one ratio, and the consideration that ions would be unlikely to occupy to neighbouring binding sites due to electrostatic repulsion, the crystallographic data were interpreted as a superposition between KWKW and WKWK occupancies at comparable populations. It was noted however, that the estimated thallium occupancy of 2.5-3 ions for the four binding sites would imply that ions sometimes occupy adjacent binding sites. Later efforts of re-refining the same datasets and analysing high-resolution data from the MthK channel [28], as well as anomalous scattering of potassium in the engineered NaK2K channel [95] and K2P TREK-1 channel [58], consistently found ion occupancies of near unity for all four binding sites in the SF in its conductive conformation, implying a vanishingly small water occupancy in the filter and a high fraction of direct ion-ion contacts. A very recent, crystallographic dataset for the MthK channel suggested a slightly lower occupancy of about 3.2 ions in the SF (with the lowest occupancy for the S2 binding site), similar to the number of Tl$^+$ ions originally reported for KcsA [96]. Although these crystallographic studies were carried out under cryogenic conditions (low temperature), and in the absence of an electrochemical gradient - hence with a channel that does not carry a net current - the filter occupancy under these conditions still represents low-energy states that can be assumed to be informative on the configurations involved in ion permeation. These high-resolution analyses of multiple potassium channels show a high occupation with ions, approaching full ion occupancy. This in turn indicates a high prevalence of direct ion-ion contacts, and therefore the recent crystallographic analyses are more consistent with the direct knock-on permeation mechanism than with soft knock-on.

This picture was recently challenged by a crystallographic study of a selectivity filter mutant of KcsA [97]. In this study, the conserved glycine (G77) contributing with its carbonyl group to the S2 and S1 ion binding sites was mutated to alanine and cysteine, respectively (G77A, G77C). In the crystal structures, the mutant proteins were shown to bind ions at the S2 and S4 binding sites. As this is one of the configurations (namely WKWK or [S2, S4]) proposed to predominate in the soft knock-on mechanism, this was taken as evidence to support soft knock-on, together with the previously known T75C (threonine forming the S4 ion binding site) mutant [98], that seems to prefer ions bound to sites S1 and S3 (i.e. KWKW, [S1, S3]). However, the low single channel conductance of these mutants, as well as the reduced $K^+/Na^+$ selectivity of the mutant equivalent to G77C in the *Shaker* $K^+$ channel [15], put into question the assumption of an unchanged permeation mechanism as compared to wild-type channels. The T75A mutant in KcsA seems to retain its $K^+$ selectivity [99], but the equivalent mutation in both MthK and NaK2K yields non-selective channels [8], thus calling for additional electrophysiological, structural and computational characterization. Furthermore, the difference in ion occupancy in structures of the G77A and G77C KcsA mutants, as compared to the wild type channels showing full ion occupancy, would need to be reconciled in order to support the assumption of an unaltered permeation mechanism in these mutants. It thus remains unclear if these mutated channels are informative on the ion occupancy and permeation mechanism of canonical, wild type potassium channels.

*Cryo-EM*

In principle, structural determination of a potassium channel with cryo-electron microscopy (cryoEM) should provide direct insights into the occupancy of the SF through ion densities, complementary to those obtained through X-ray crystallography, with the additional advantage of retaining the membrane environment around the channel. Indeed, structures of multiple members of the potassium channel family have been obtained in lipid nanodiscs, including voltage gated human channels (hERG, Kv11.1) [7], $Ca^{2+}$ and $Na^+$ gated big conductance (BK) channels [100], the voltage gated Kv1.2-2.1 paddle chimera channel [101], the voltage gated KAT1 channel [102], the voltage and cyclic-nucleotide gated SthK channel [103], the voltage gated Eag channel (Kv10.1) [104] and the pH gated K2P TASK2 channel [102]. However, none of these structures reached a resolution higher than 3.0 Å, especially in the transmembrane part (where the SF is located), and therefore, although the SF is seen in the conductive conformation in many of these structures, indistinguishable from x-ray structures, an equivocal determination of the SF occupancy is precluded. We expect, however, that with the recent breakthroughs in cryo-EM, which enable atomic level resolution for both soluble and membrane proteins [105,106], a glimpse into the SF content will be available in the near future.

*Nuclear Magnetic Resonance (NMR)*

NMR spectroscopy offers an appealing experimental way to study the dynamics of proteins in near-physiological environments and temperatures, although the membrane voltage is not taken into account thus far. For membrane proteins, both solution NMR can be used, where the protein is typically solvated in detergent micelles, e.g. dodecyl maltoside (DDM), n-dodecylphosphocholine (DPC), or detergent/lipid mixed micelles, as well as solid-state NMR (ssNMR), that allows protein samples to be embedded in detergent-free liposomes and bicelles made of lipids [107]. In the context of potassium channels, many NMR experiments have been published, focusing mostly on the gating processes that occur both at the main activation gate and at the selectivity filter [108–110], as well as on the coupling between these two gates. Even though NMR is a well suited technique to probe the occupancy of the selectivity filter, due to the high sensitivity of chemical shifts to the local environment, most of the earlier studies assumed that the occupancy states characteristic for the knock-on mechanism (KWKW and WKWK), or the occupancy of the SF was not discussed at all. Already in 2010 however, Imai *et al.* observed, via analysis of Nuclear Overhauser Effect (NOE) patterns, that the open to inactivated channel transition in KcsA involves binding of a water molecule to the filter and elimination of (some) $K^+$ ions, in agreement with crystallographic data (see Figure 1 D) [111]. The hypothetical presence of water in the filter in the open, conductive state, as required by the knock-on mechanism, was then explained by a possible residence time of such waters in the studied site (V76 from the SF) being shorter than 300 ps, and thus not detectable by NOE.

The proposal of the direct knock-on mechanism led to a subsequent design of NMR experiments to specifically address the occupancy states of the SF. In 2019, Öster et al. used ssNMR to probe the interactions of the engineered potassium channel NaK2K with water molecules [56]. Specifically, hydrogen/deuterium (H/D) exchange and 1H (amide protons) spin diffusion, via magnetization exchange, experiments were performed. The NaK2K channel presents an additional advantage for such investigations, as its SF remains in the conductive conformation at a wide range of conditions, including low potassium concentration [26], at which filters of some other channels, e.g. KcsA and Kv channels, collapse/inactivate (Figure 1), possibly obscuring the interpretation of an experiment. The H/D exchange pattern for NaK2K showed signals only for the upper part of the SF (G67 and Y66 in NaK2K, forming ion binding sites S1 and S2). The signal from G65 was weak, and signals from the lower part of the SF (V64 and T63) were absent. This observation indicates that the ion binding sites S4 and S3 are not in contact with water. As a control, removing all potassium ions from the sample led to clear H/D signals from V64 and T63, which are thus exposed to water in potassium-free conditions (i.e. water molecules occupy the SF when no ions are available). Re-introduction of potassium ions into the sample caused these signals to disappear, suggesting that K$^+$ ions are indeed replacing all water molecules, at least in ion binding sites S3 and S4. In a second experiment, magnetization transfer from water to protein was used to provide another way of measuring which residues are in contact with water, importantly on a shorter time scale of 1 to 100 ms. The efficiency of the transfer depends on the distance between the bound water molecules and the protein (amide protons). The residues from the SF that showed magnetization transfer were GYG (65-67) (Figure 1). These residues are in contact with two structural water molecules (in the NaK2K channel), necessary for filter stability [112,113]. Indeed, in all high resolution structures of potassium channels, such water molecules are resolved (Figure 3) and their presence stabilizes the SF in the conductive conformation (see the discussion in the computational part as well). Accordingly, in the ssNMR experiment, the signal buildup behavior for magnetization transfer was followed (based on different experimental transfer times). The resulting signal buildup rate shows a clear dependence on the distance to the structural waters bound behind the SF, that is the rate is the highest for G67 (top of the SF), followed by Y66 and lowest for G65 (middle of the SF), and not detectable for either V64 nor T63. Consequently, if there were any water molecules bound in the filter (i.e. in the ion binding sites, as postulated by the soft knock-on mechanism), they should affect the magnetization transfer buildup rate. These two experiments, together with MD simulations carried out in the absence of membrane voltage, led to the conclusion that the SF of potassium channels, in its conductive conformation, is water-free under near-physiological conditions (i.e. in a lipid membrane and at room temperature). Such a conclusion strongly supports the direct knock-on mechanism, although these ssNMR experiments did not address the actual occupancy of the filter by $K^+$ ions, and how many $K^+$ ions are present in the filter at a given time (instantaneous occupancy). Even more recently, a similar lack of water accessibility to the inner ion binding sites has been observed, also via ssNMR, in the inward-rectifier KirBac1.1 channel [114].

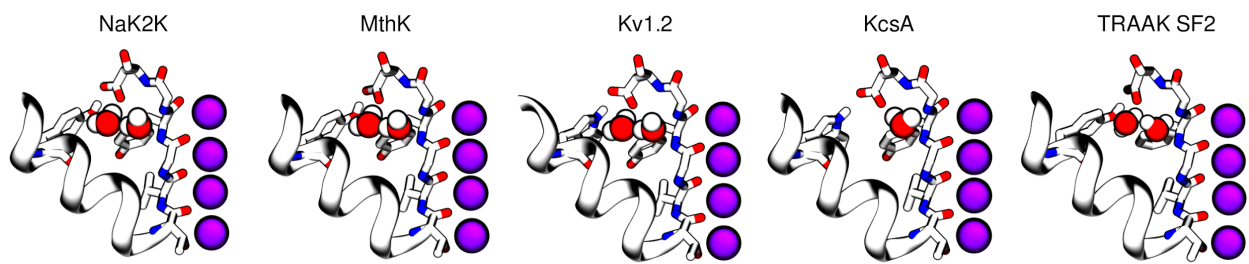

*Figure 3. Structural water molecules (red and white spheres) bound behind the selectivity filter. In high resolution structures of several potassium channels, for example NaK2K (PDB ID: 6UFE), MthK (PDB ID: 3LDC), Kv1.2 (PDB ID: 2R9R), KcsA (PDB ID: 1K4C) and TRAAK (PDB ID: 4I9W), water molecules are resolved behind the SF, taking part in an intricate hydrogen bond network. In MD simulations, these water molecules exchange with bulk waters slowly (hundreds of ns) and help to stabilize the SF in the conductive conformation.*

In another study, Eichmann *et al.* used solution-state NMR to study the binding of $NH_4^+$ ions to the KcsA channel, embedded in detergent micelles [115]. $NH_4^+$ ions are considered as good surrogates of $K^+$ ions, both from the physicochemical perspective (ionic radii differ only by ~0.15 Å and both ions have comparable hydration free energies [78]), as well as from their ability to permeate potassium channels [5], suggesting a similar mode of binding to the SF and the mechanism of ion permeation shared by these two ions (although currents carried by $NH_4^+$ are typically markedly lower than $K^+$ currents). Crucially, $NH_4^+$ ions containing the $^{15}N$ isotope provide high sensitivity for NMR investigations. The NMR spectra revealed 5 signals indicative of $NH_4^+$ ions bound to the channel (ascribed to the ion binding sites S0 - S4). The signals disappeared after addition of K$^+$ ions to the sample, suggesting a competition between K$^+$ and NH$_4^+$ ions for the same ion binding sites, most likely at the SF. The disappearance of signals in a control experiment with the scorpion toxin agitoxin II, which plugs the SF and displaces ions from it [116], reinforced the idea that the observed NH$_4^+$ signals indeed stem from the ammonium ions bound to the SF. By integrating $^{15}$NH$^+$ cross peaks and comparing their volumes with cross peak volumes of $^{15}$N-$^1$H (amide) moieties, the authors were able to estimate the number of NH$_4^+$ ions in the SF at a given time to be in the range of 3-4, strongly suggesting the existence of direct ion-ion contacts between the ions and a lack of water molecules in the filter at zero voltage and a temperature of 36ºC.

*Functional measurements*

Early experimental work by Hodgkin and Keynes in 1955 determining the ion flux-ratio of K$^+$ channels arrived at the result that K$^+$ ions do not traverse the channels independently, but that about three ions moved "in lockstep" with each other [11]. Flux-ratio measurements performed later with technology of increased precision [117] recorded the flux-ratio exponent of Shaker K$^+$ channels as 3.4 and interpreted this result as indicating that "the pore in these channels can simultaneously accommodate at least four K+ ions" [117,118]. In general, the analysis of flux-ratio measurements is not straightforward: depending on the model used, the value obtained predicts only the lower boundary of the number of ions in the pore [117]. Moreover, ion concentrations can significantly affect this value [119]. However, in [117] the measurements were performed at a low extracellular concentration of K$^+$ (25 mM), far from the concentration at which the current saturates. Under these conditions, the flux-ratio is equal to the number of ions in the pore [120,121]. These early measurements thus suggest high ion occupancy of the SF, in agreement with the direct knock-on mechanism.

The most direct, albeit difficult to interpret, experiments regarding the ratio of water to ion flux during $K^+$ channel permeation are streaming potential measurements. Streaming potentials are recorded when ion channels experience a strong osmotic gradient from one side of the membrane to the other under otherwise symmetrical ionic concentrations. The osmotic pressure drives water through the pore, which sweeps ions from the ionic solution with it, building up a small voltage difference across the membrane as a consequence. The literature cites a small number of key papers, which determine streaming potentials in the $Ca^{2+}$ coupled large conductance $K^+$ channel [122], the $K^+$ channel from sarcoplasmic reticulum [123], the human Ether-a-go-go-Related channel (hERG) [124], and most recently, KcsA [125]. Due to the small magnitude of the streaming potentials, the measurements are however difficult to carry out, rather noisy, and the results are often described as hard to interpret, partly due to the necessary use of correction factors in the conversion to water-ion flux ratios. For instance, potassium permeation mediated by the selective membrane carrier valinomycin generates a measurable streaming potential, although it is assumed not to be associated with water co-translocation [122]. The valinomycin signal is hence assumed to arise from the direct effect of the osmotic gradient on the ion activity coefficient, and not as part of the actual streaming potential. Thus, valinomycin control experiments are frequently carried out as control to derive the channel related streaming potential after correcting for this effect.

The original papers report that per $K^+$ ion, between 2 and 4 water molecules co-permeate in the $Ca^{2+}$ coupled $K^+$ channel, inferred from a (corrected) streaming potential of 1-2 mV [122]. For the $K^+$ channel from sarcoplasmic reticulum, the water-to-ion ratio was found to lie between 2-3 [123]. In 2005, Ando et al. introduced a new approach to record streaming potentials using an osmotic-pulse methodology [124]. In hERG, the authors derived a water-ion flux ratio between 0.9-1.4, depending on $K^+$ concentration, while for KcsA, a ratio between 1.0-2.2 emerged, where the higher numbers are associated with more dilute ionic solutions. Remarkably, the upper limit in KcsA is obtained from recording streaming potentials due to ion flux at 3mM [K$^+$], a concentration at which the channel typically inactivates and becomes non-conductive to K$^+$ ions [20].

In further experiments, Pohl and colleagues recorded water flux through KcsA channels by determining the dilution of ionic solutions by inflowing water near the membrane [126], and by stopped-flow measurements of the volume of proteoliposomes under osmotic gradients [127]. In both studies, the thus deduced diffusion coefficient of water inside KcsA exceeded that of bulk water at low K$^+$ concentrations, whereas K$^+$ concentrations above 300 mM were found to completely block water permeation through the channels. Inactivated KcsA channels, impermeable to K$^+$ ions, were still observed to allow water passage at similar rates, while open channels under 200 mM K$^+$ displayed an intermediate osmotic permeability in "qualitative agreement with single file movement of K$^+$ and water" [104].

A molecular mechanistic interpretation of these findings is not straightforward. Firstly, the results imply that water flux through KcsA is 100 times faster than K$^+$ permeation at low [K$^+$], and that at 200 mM K$^+$, the rate of water permeation still far exceeds that of K+ conduction, which is difficult to reconcile with a 1:1 co-permeation mechanism. Secondly, raised K$^+$ levels reduce water flux to a non-detectable level, leaving open the possibility of K$^+$ permeation without water. The simplest explanation that might resolve this puzzle is to assume that water flow occurs through a subset of KcsA channels unoccupied with K$^+$, since most channels do not contain an ion at the given experimental conditions [127] - whereas open, active channels filled with K$^+$ ions at high concentration are less or indeed non-permeable to water.

Rauh *et al.* [128] used an electrophysiological approach in a study that originally aimed to characterize unusual voltage gating in viral potassium channels. In contrast to classical voltage sensing and gating, which is mediated by voltage-sensing domains (VSDs), some potassium channels show a distinct voltage gated process, which is thought to be mediated by changes of $K^+$ ion occupancy in the selectivity filter. As both ion permeation and such voltage gating occur at the SF and involve $K^+$ ions moving between different sites, it was assumed that atomistic models of ion permeation would provide insights into voltage gating at the SF as well. The authors used single-channel data, namely IV curves, to test three different mechanisms of ion permeation, two of which resembled the soft knock-on mechanism (no simultaneous occupancy of adjacent binding sites by $K^+$ in the SF), whereas the third mechanism involved some states characteristic for the direct knock-on mechanism. The considered mechanisms varied in the number of discrete states and rate constants between them and in the ion occupancy of such states and the voltage (in)dependence of certain transitions. The mechanisms were subsequently tested by a global fit of 36 experimentally derived curves obtained at varying $K^+$ concentration and membrane voltages. The only mechanism that fit all the electrophysiological data was one of the variants of the knock-on mechanism, the so-called '5 state model', therefore supporting the original knock-on mechanism. It is however important to note that the mechanism related to the direct knock-on considered by Rauh *et al.* was only one possible variant (see Figure 2). The approach did not provide any information regarding the presence of water in the SF, nor the instantaneous number of $K^+$ ions in the filter during permeation, which is expected to be close to 2 for the knock-on mechanism, but higher for direct knock-on.

In contrast, Schewe *et al.* [57] studied a similar type of voltage gating in the SFs of K2P channels, also with electrophysiological methods. Here, the authors were able to measure the magnitude of gating charges - movements of charged moieties within the membrane that change their positions during gating. As K2P channels do not possess VSDs, which contain charged residues as gating charges, it was postulated that $K^+$ ions (and their surrogates, i.e. $Tl^+$, $Rb^+$ and $Cs^+$) moving in and out of the SF must act as gating charges. The measurements with $Rb^+$ and $Cs^+$ ions revealed that 3-4 ions have to enter the closed/collapsed SF from Scav to allow the conformational transition to the open state. This estimation requires an assumption that the most (ca. 80%) of the membrane electric field is focused on the filter, between S1 and S4, and that Scav lies outside of it. In such a scenario, these three to four ions would need to occupy three to five ion binding sites (S1-S3, plus possibly S0 and S4). However, as S4 might be occupied also in inactivated filters [58], and S0 lies outside of the assumed membrane electric field, the contributions from these sites to the measured gating charge would be lower. Consequently, almost all possibilities of accommodating such a higher number of ions in the SF would require formation of direct ion-ion contacts, characteristic for the direct knock-on mechanism.

*Two-dimensional infrared spectroscopy (2D IR)*

Another experimental technique proposed to study the ion permeation mechanism in potassium channels is two-dimensional infrared spectroscopy (2D IR), which, in principle, allows studying the SF in a membrane-like environment with native $K^+$ ions and at zero voltage. The chemical bond vibrations are sensitive to the electrostatic environment, and therefore the vibration frequencies will be markedly different depending on the presence of ions and/or water nearby [129]. Of particular interest here is the amide I stretching mode, dominated by the stretching motion of the carbonyl group. As the ion binding sites S1-S3 are formed exclusively by carbonyl oxygens, the 2D IR technique seems to be well suited to experimentally probe the occupancy and dynamics of the SF of potassium channels. However, to isolate the stretching motion of the specific carbonyl groups, isotope labeling ($^{13}C$ and $^{18}O$) is necessary, which redshifts the amide I frequency. Moreover, such a 2D IR spectrum reports on an ensemble average of ion occupancy states in the SF. To interpret the spectrum, i. e. to decompose it into the contributions from specific occupancy states, MD-based theoretical spectra are used: when a theoretical spectrum, or a combination thereof, agrees with the experimental spectrum, the occupancy state(s) used to calculate the theoretical spectra are deemed as compatible with the one present under experimental conditions.

In 2016, Kratochvil *et al.* used for the first time, the 2D IR approach together with the semisynthesis of the isotopically-labeled KcsA channel and MD-derived spectra, to probe the occupancy of the SF of KcsA [130]. It was found that only spectra that were calculated from simulations of water-containing states of the SF, namely soft knock-on signature states KWKW and WKWK (Figure 2) with the addition of the "carbonyl flipped" state of the S3 valine (Figure 1 D), were in agreement with the experimental spectrum, after fitting the relative population of each state guided by the experimental spectrum. In contrast, the populations of spectra calculated from simulations of states characteristic for the direct knock-on (e.g. 0KK0 and KK0K) were found to match the experimental spectra less well. However, unlike the snapshots representing the soft knock-on, the states representing the direct knock-on were not fitted to the experimental data, but were rather taken directly from computational electrophysiology simulations by Köpfer *et al.*, i.e. at non-zero voltage. As mentioned, the 2D IR experiment is performed at zero voltage. Nevertheless, Kratochivil *et al.* concluded that only the water containing soft knock-on states KWKW and WKWK (and presumably the "carbonyl flipped" conformation) are in line with the 2D IR spectra.

We have later repeated the calculations but used the same procedure of fitting of MD-derived spectra for both mechanisms (soft and direct knock-on) [52] following the same protocol as the one used in Kratochvil *et al.* For the direct knock-on mechanism, we have also included a larger number of the SF occupancy states, to reflect those commonly seen in our simulations (e.g.

WKK0 and WKKK). In our analysis, we were able to reproduce the theoretical spectra of Kratochvil *et al.* for the soft knock-on mechanism, indicating the robustness of the method. In contrast however, when the direct knock-on states were treated with the same fitting protocol, also these yielded a very good agreement with the experimental spectra, in fact as good as the soft knock-on states. This observation led us to conclude that such an analysis, based on the experimental spectra from Kratochvil *et al.*, cannot differentiate between the occupancy states characteristic for soft and direct knock-on mechanisms, and thus could not discriminate between both mechanisms of ion permeation.

It is worth noting that the experimental spectra for the labelled KcsA discussed here come from one specific isotope-labelling pattern, namely $^{13}C^{18}O$ labels simultaneously introduced on the carbonyl groups of V76 (S3 binding site), G77 (S2 binding site), G79 (S1 binding site) (see Figure 1 C). The simultaneous IR signal from all three labelled residues can obstruct the interpretation of the experimental spectra. It has been recently argued, based on theoretical calculations, that isotope labeling of single residues, namely V76 and G77, in independent measurements, should be sensitive to the water presence and/or direct contacts between $K^+$ ions in sites S2 and S3, thus distinguishing between the soft and direct knock-on mechanisms [131]. However, as of now (early 2021), the corresponding experimental spectra have not been published.

**Concluding remarks**

The ion permeation mechanism in potassium channels has intensely been studied in the past decades by an impressive number of experimental, computational, and theoretical approaches. Yet, as reviewed above, there is no agreement on what permeation mechanism is dominant, especially at physiological conditions. For convenience, we summarized the outcomes of the studies reviewed here in Table 1.

In our opinion, the main open questions are: i) how many ions occupy the selectivity filter of potassium channels at any given time? ii) what are the essential steps during ion permeation? iii) does water co-permeate with potassium ions under physiological conditions, and, if yes, to what extent? iv) can potassium ions occupy neighboring ion binding sites? v) are currently available molecular force fields accurate enough to simulate ionic currents through potassium channels quantitatively in computational studies? If not, what improvements are critical?

Possible reconciling scenarios would include the possibility that under different conditions, such as channel composition or ion concentration, a different mechanism prevails, or that particular experiments primarily "see" a particular filter conformation or sub-population, which may or may not be representative for the major ion conductive state.

Table 1. Agreement between various computational and experimental studies and proposed ion permeation mechanisms in potassium channels. Name of the channel for which the given observation was recorded is given in parentheses. ✔ means that the main observations are compatible with a given mechanism, whereas ✗ means it would be very hard to reconcile the two. **?** stands for observations that can be interpreted differently or potential issues with the employed method. See text for the full discussion.

| Observable | Compatible with... | |
|---|---|---|
| | **knock-on** | **direct knock-on** |
| **Absolute occupancy of ion binding sites by X-ray crystallography [KcsA, MthK, NaK2K]** | ✗ | ✔ |
| **Water presence in SF by ssNMR [NaK2K]** | ✗ | ✔ |
| **Number of ions ($NH_4^+$) in the SF by NMR [KcsA]** | ✗ | ✔ |
| Gating charge by electrophysiology [several K2P channels] | ✗ | ✔ |
| Number of ions and water molecules in the SF by X-ray crystallography [KcsA mutants] | ✔ | ?/ ✗ |
| Flux-ratios [Shaker] | ✗ | ✔ |
| SF occupancy by 2D IR spectroscopy [KcsA] | ✔ | ✔ |
| Streaming potentials [KcsA, hERG, BK] | ✔ | ?/ ✗ |
| Simulated currents by MD [KcsA, MthK, NaK2K, Kv1.2, TRAAK, TREK-2] | ?/ ✗ | ✔ |
| Simulated $K^+$/$Na^+$ selectivity [KcsA, MthK, NaK2K, Kv1.2] | ?/ ✗ | ✔ |
| Global fits by electrophysiology [$Kcv_{NTS}$] | ✔ | ✗ |

## ACKNOWLEDGEMENTS

This work was supported by the German Research Foundation DFG through FOR 2518 DynIon, Project P5 (A. M., B. d. G., W. K). We thank Daniel L. Minor, Andrew Natale, Michael Grabe and Andrew Plested for many useful comments and Petra Kellers for editorial assistance.